# Effect of new healthy and live food supplement on Anemia disease in Wistar rats

Azam Bayat [1, 2, *], Aref Khalkhali [2], Ali Reza Mahjoub[1]

[1]*Department of Chemistry, Tarbiat Modares University, Tehran 14155-4383, Iran*

[1]*NBS organic Company, Istanbul, Turkey*

**Corresponding author: Azam Bayat**

**E-mail address: azam.bayat@modares.ac.ir**

**Abstract**

Anemia is a decrease in hemoglobin and red blood cells and due to a decrease in hemoglobin, oxygen carrying capacity reduce. In this disease, the red blood cell the amount and volume decrease. In this research, healthy and live food powder were synthesized by a green route. This organic biomaterial was named NBS. The NBS healthy and live food powder has various vitamins, macro and micro molecules, and ingredients. Twenty Wistar rats were randomly divided into 4 equal groups, including control and treatment groups 1, 2 and 3. Nutritional supplements for healthy living were administered orally via gavage to rats in groups 1, 2, and 3 at 12.5, 25, and 50 mg/ kg, respectively, and within a period of 20 days, one day in between. There was no intervention in the control group in order to reach baseline blood factors. At the end of the study, blood samples were taken from the heart, including blood-red blood cells, hemoglobin, hematocrit and platelets using a fully automated blood cell counting machine. The results showed that the new dietary supplement reduced the level of hematocrit and platelets in the studied rats. The healthy and live food supplement at a concentration of 50 mg / kg increased blood levels compared to the control group. The results of this study showed that the use of healthy and live food supplement increased blood factors compared to the control group.

## Introduction

Anemia is a decrease in hemoglobin and red blood cells and due to a decrease in hemoglobin, oxygen carrying capacity reduce. In this disease, the red blood cell the amount and volume decrease. Anemia has many different types, each of which occurs for different reasons, briefly as follows [1]:

**1. Megaloblastic anemia**

This type of anemia is usually caused by folic acid or vitamin B12 deficiency. These vitamins help keep the blood or nervous system healthy. In this type of anemia, the body produces red blood cells that cannot deliver oxygen well. If this type of anemia is due to vitamin B12 deficiency, it can cause numbness in the hands and feet, difficulty walking, memory loss, and vision problems. The type of treatment depends on the causative agent, but it is nevertheless necessary to consume vitamin B12 [2].

**2. A latent illness**

There are certain diseases that can impair the body's ability to produce red blood cells. For example, people with kidney disease, especially dialysis patients, are at risk for anemia. These kidneys are unable to produce enough hormone (erythropoietin) to produce blood cells and also lose iron during dialysis [3].

**3. Hereditary anemia**

If a family has a history of blood disease, it is also more likely to develop in other people. One of the hereditary blood diseases is sickle cell type anemia. In this type, instead of producing natural red blood cells that move easily through the blood vessels, sickle cells are made that are hard to move and have curved edges. These cells cannot move easily into the fine blood vessels. And thus block the way blood reaches the organs. The body destroys the sickle cells. But it cannot produce new natural and healthy varieties quickly enough. This causes hemorrhage [4, 5].

Another type of hereditary anemia is thalassemia. Thalassemia occurs when the body loses certain genes or the abnormal genes are inherited from parents to the baby, which have a negative impact on the production of hemoglobin [6].

**4. Aplastic anemia**

It is a rare form of anemia and occurs when the body does not produce enough red blood cells. Because it affects white blood cells, the risk of infections and non-stop bleeding increases. There are several reasons for this [7-8]:

a. Cancer Therapies (Radiation and Chemotherapy)

B. Exposure to toxic chemicals (such as substances used in some insecticides, paints, cleaners and household detergents)

P. Some medications (such as rheumatoid arthritis)

D. Autoimmune diseases (such as lupus)

E. Viral infections that affect bone density.

**5. Iron Deficiency Anemia (IDA)**

If iron is not available enough to make red blood cells, the body uses its reserve first, and if the deficiency continues, the body's iron stores decrease, and when iron deficiency continues, the body's stores are depleted and Iron deficiency anemia occurs [9].

**I. Causes of Iron Deficiency Anemia**

1. Blood loss occurs due to problems such as long and heavy monthly habits, internal wounds, colon polyps and colon cancer.

2. Diets without iron.

3. Iron absorption disorder due to diarrhea, decreased gastric acid secretion, gastrointestinal problems or drug interactions such as cholestyramine, cimetidine, pancreatin, ranitidine and tetracycline.

4. Intestinal parasites, especially hookworms

5. Rapid growth in infancy and adolescence and increased need in pregnancy and lactation

## II. Symptoms of Iron Deficiency Anemia

Painful skin, tongue and mucus in the lips and eyelids, early fatigue, dizziness and headache, sleepiness and tingling in the legs, nausea, anorexia and severe anemia (nail spasm), palpitations and shortness of breath and in women have been noted as excessive and prolonged haemorrhage in the habit of monthly or reduced volume of bleeding and even cessation of bleeding during this period.

Iron deficiency can lead to delayed physical development, impaired brain development, impaired children's intelligence, decreased learning ability, and academic failure. Also, the IQ of children with iron deficiency is 5 to 10 points lower than non-anemic children. The effects of anemia in infancy and early childhood are irreversible and cannot be remedied by subsequent therapies [7, 8].

Iron deficiency reduces the body's resistance to infection. And a person with iron deficiency is more likely to have the disease and have a longer period of illness. In adults with iron deficiency, the ability to concentrate is low. These people are not capable of doing their daily tasks and soon become weak and tired. Also, women with iron deficiency are often bored, unwilling to work, and soon feel tired, which in turn reduces their level of family care.

Pregnant mother's anemia due to iron deficiency can delay fetal growth, low birth weight (less than 2500 g) birth, and increase mortality around birth. Losing blood during childbirth is very dangerous for a mother who has severe anemia and can cause her death.

However, these symptoms are not very specific and accurate, and the diagnosis of iron deficiency is largely based on tests [10].

### III. Prevalence of Iron Deficiency Anemia

Based on available information and the amount of hemoglobin recommended by the WHO Scientific Society, an estimated 30% of the world's population has anemia. Children and pregnant women are more vulnerable than other groups and the overall prevalence of anemia is 43% and 51%,respectively.

In our country, a national survey of micronutrients status by the National Bureau of Nutrition and Food Improvement and the National Reference Laboratory in 2001 showed that about 40% of children aged 15-23 months, 18% of children aged 6 years and 20 years 14-year-olds and 21% of pregnant women in the country's 11 climates have anemia based on hemoglobin. Iron deficiency based on serum ferritin is present in about 33% of 15-23 month olds, 26% of 6 year olds and 23% of girls and boys as well as 43% of pregnant women in the country [11-12].

In this research, healthy and live food powder were synthesized by a green route. This organic biomaterial was named NBS. The NBS healthy and live food powder has various vitamins, macro and micro molecules, and ingredients. The results of this study showed that the use of healthy and live food supplement increased blood factors compared to the control group.

## Materials and Methods

Twenty Wistar rats were selected in a weight range of 220 ± 0.5 g and maintained at temperatures of 25 to 30 ° C and free access to food and water. The mice were randomly divided into four groups, including control and treatment groups 1, 2 and 3 respectively. Nutritional supplements for healthy living were administered orally via gavage to rats in groups 1, 2, and 3 at 12.5, 25, and 50 mg / kg, respectively, and within a period of 20 days, one day in between. There was no intervention in the control group in order to reach baseline blood factors. At the end of the study, blood samples were taken from the heart, including blood-red blood cells, hemoglobin, hematocrit and platelets using a fully automated blood cell counting machine.

## Results and Discussion

Analysis of phenolic compounds in the Nutritional supplement and healthy living (extracted with using HPLC) are shown as Fig. 1. The percent of phenolic compounds of new healthy and live food supplement are Arctigenin 2.34, Gallic Acid 2.41, Quercetin 9.42, Alpha Linoleic Acid 26.80, Linoleic Acid 19.46, Inulin 2.64, Oleic acid 13.24, Palmitic acid 14.98, Stearic acid 3.14 and unknown  compound.

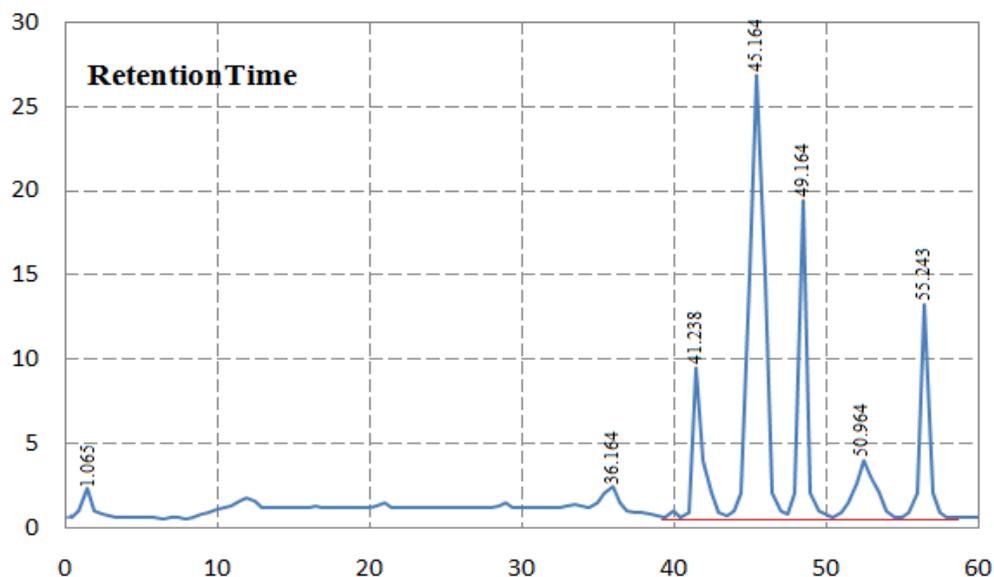

**Fig. 1.** Analysis of phenolic compounds in the Nutritional supplement and healthy living (extracted with using HPLC)

The results of this study showed that the use of a healthy diet resulted in a significant increase in erythrocyte and hemoglobin concentrations in the blood of the mice studied, while the use of this diet reduced the hematocrit and Platelets in studied rats (Table S1-S4 in supplementary materials). In the study of the interactions of the studied concentrations, it can be concluded that the use of 50 mg/ kg dietary supplement increases blood levels compared to the control group (Fig. 2).

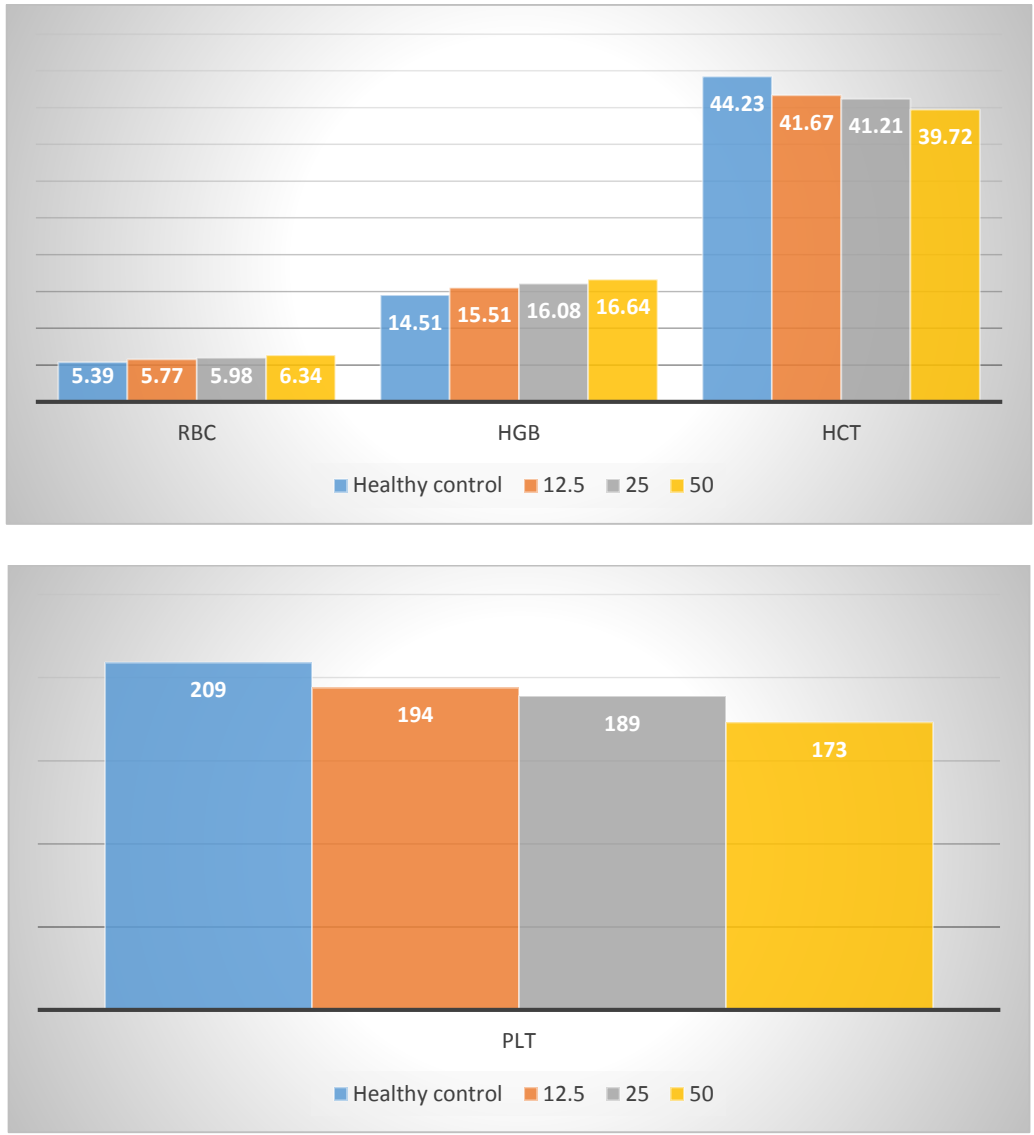

**Fig. 2.** Comparison of blood factors in control mice and treated with a food supplement and live healthy

In this study, in order to evaluate the significance of the data, it is recommended to use ANOVA test. ANOVA test was used to investigate the differences between and within groups. The results of this test showed that there was a significant difference between the groups with 5% probability level. In order to clarify this issue, by Scheffe post hoc test, this significance was tested one by one between groups. The results can be seen in Table 1-4.

| (I) VAR00001 | (J) VAR00001 | Mean Difference (I-J) | Std. Error | Sig. | 95% Confidence Interval | |
|---|---|---|---|---|---|---|
| | | | | | Lower Bound | Upper Bound |
| control group | Concentration 12.5 | -.51400* | .05393 | .000 | -.6821 | -.3459 |
| | Concentration 25 | -.66200* | .05393 | .000 | -.8301 | -.4939 |
| | Concentration 50 | -.90000* | .05393 | .000 | -1.0681 | -.7319 |
| Concentration 12.5 | control group | .51400* | .05393 | .000 | .3459 | .6821 |
| | Concentration 25 | -.14800 | .05393 | .096 | -.3161 | .0201 |
| | Concentration 50 | -.38600* | .05393 | .000 | -.5541 | -.2179 |
| Concentration 25 | control group | .66200* | .05393 | .000 | .4939 | .8301 |
| | Concentration 12.5 | .14800 | .05393 | .096 | -.0201 | .3161 |
| | Concentration 50 | -.23800* | .05393 | .004 | -.4061 | -.0699 |
| Concentration 50 | control group | .90000* | .05393 | .000 | .7319 | 1.0681 |
| | Concentration 12.5 | .38600* | .05393 | .000 | .2179 | .5541 |
| | Concentration 25 | .23800* | .05393 | .004 | .0699 | .4061 |

*. The mean difference is significant at the 0.05 level.

**Table 1.** Results of ANOVA analysis and Scheffe post hoc test data for RBC results of rats studied in this study

| (I) VAR00001 | (J) VAR00001 | Mean Difference (I-J) | Std. Error | Sig. | 95% Confidence Interval | |
|---|---|---|---|---|---|---|
| | | | | | Lower Bound | Upper Bound |
| control group | Concentration 12.5 | -1.13200* | .21410 | .001 | -1.7994 | -.4646 |
| | Concentration 25 | -1.79400* | .21410 | .000 | -2.4614 | -1.1266 |
| | Concentration 50 | -2.15800* | .21410 | .000 | -2.8254 | -1.4906 |
| Concentration 12.5 | control group | 1.13200* | .21410 | .001 | .4646 | 1.7994 |
| | Concentration 25 | -.66200 | .21410 | .052 | -1.3294 | .0054 |
| | Concentration 50 | -1.02600* | .21410 | .002 | -1.6934 | -.3586 |
| Concentration 25 | control group | 1.79400* | .21410 | .000 | 1.1266 | 2.4614 |
| | Concentration 12.5 | .66200 | .21410 | .052 | -.0054 | 1.3294 |
| | Concentration 50 | -.36400 | .21410 | .434 | -1.0314 | .3034 |
| Concentration 50 | control group | 2.15800* | .21410 | .000 | 1.4906 | 2.8254 |
| | Concentration 12.5 | 1.02600* | .21410 | .002 | .3586 | 1.6934 |
| | Concentration 25 | .36400 | .21410 | .434 | -.3034 | 1.0314 |

*. The mean difference is significant at the 0.05 level.

**Table 2. .** Results of ANOVA analysis and Scheffe post hoc test data for HGB results of rats studied in this study

| (I) VAR00001 | (J) VAR00001 | Mean Difference (I-J) | Std. Error | Sig. | 95% Confidence Interval | |
|---|---|---|---|---|---|---|
| | | | | | Lower Bound | Upper Bound |
| control group | Concentration 12.5 | 1.50600* | .23520 | .000 | .7728 | 2.2392 |
| | Concentration 25 | 2.78600* | .23520 | .000 | 2.0528 | 3.5192 |
| | Concentration 50 | 3.81600* | .23520 | .000 | 3.0828 | 4.5492 |
| Concentration 12.5 | control group | -1.50600* | .23520 | .000 | -2.2392 | -.7728 |
| | Concentration 25 | 1.28000* | .23520 | .001 | .5468 | 2.0132 |
| | Concentration 50 | 2.31000* | .23520 | .000 | 1.5768 | 3.0432 |
| Concentration 25 | control group | -2.78600* | .23520 | .000 | -3.5192 | -2.0528 |
| | Concentration 12.5 | -1.28000* | .23520 | .001 | -2.0132 | -.5468 |
| | Concentration 50 | 1.03000* | .23520 | .005 | .2968 | 1.7632 |
| Concentration 50 | control group | -3.81600* | .23520 | .000 | -4.5492 | -3.0828 |
| | Concentration 12.5 | -2.31000* | .23520 | .000 | -3.0432 | -1.5768 |
| | Concentration 25 | -1.03000* | .23520 | .005 | -1.7632 | -.2968 |

*. The mean difference is significant at the 0.05 level.

**Table 3.** Results of ANOVA analysis and Scheffe post hoc test data for HCT results of rats studied in this study

| (I) VAR00001 | (J) VAR00001 | Mean Difference (I-J) | Std. Error | Sig. | 95% Confidence Interval | |
|---|---|---|---|---|---|---|
| | | | | | Lower Bound | Upper Bound |
| control group | Concentration 12.5 | 13.40000* | 2.27376 | .000 | 6.3123 | 20.4877 |
| | Concentration 25 | 23.20000* | 2.27376 | .000 | 16.1123 | 30.2877 |
| | Concentration 50 | 33.60000* | 2.27376 | .000 | 26.5123 | 40.6877 |
| Concentration 12.5 | control group | -13.40000* | 2.27376 | .000 | -20.4877 | -6.3123 |
| | Concentration 25 | 9.80000* | 2.27376 | .005 | 2.7123 | 16.8877 |
| | Concentration 50 | 20.20000* | 2.27376 | .000 | 13.1123 | 27.2877 |
| Concentration 25 | control group | -23.20000* | 2.27376 | .000 | -30.2877 | -16.1123 |
| | Concentration 12.5 | -9.80000* | 2.27376 | .005 | -16.8877 | -2.7123 |
| | Concentration 50 | 10.40000* | 2.27376 | .003 | 3.3123 | 17.4877 |
| Concentration 50 | control group | -33.60000* | 2.27376 | .000 | -40.6877 | -26.5123 |
| | Concentration 12.5 | -20.20000* | 2.27376 | .000 | -27.2877 | -13.1123 |
| | Concentration 25 | -10.40000* | 2.27376 | .003 | -17.4877 | -3.3123 |

*. The mean difference is significant at the 0.05 level.

**Table 4.** Results of ANOVA analysis and Scheffe post hoc test data for PLT results of the rats studied in this study.

Anemia is a common blood disorder that does not have enough red blood cells or hemoglobin. Hemoglobin in the red blood cells causes oxygen to bind to the red blood cell and reach different tissues in the body by capillaries. Since all cells of the human body are dependent on oxygen for survival, its deficiency causes hypoxia and presents a wide range of problems. Anemia has many different types, each with its own cause. The disease is categorized on a variety of grounds, such as erythrocyte morphology, etiological mechanisms, and so on. The three main categories are: loss of large volume of blood (acute or chronic bleeding), destruction of blood cells (hemolysis), and lack of blood cell production. Blood is made up of a blood and fluid cell called a plasma. Two types of blood cells (platelets are not cells but cellular components) are floating in the plasma [11-12]. White blood cells fight infections. Platelets (Cytoplasmic fragments originating from megakaryocyte cells) help to clot blood after injury. Red blood cells: These cells carry oxygen from the lungs to the brain, various other organs and tissues of the body, your body needs oxygenated blood for its functions. Oxygenated blood helps to provide the body with the energy it needs, and it generates warmth and healthy skin color. Hemoglobin is an iron-containing protein substance that causes red blood to circulate is the key factor in the transfer of oxygen from the lungs to other parts of the body, and it also transports carbon dioxide from the tissues of the body to the lungs to be excreted through the lungs. Red blood cells are constantly produced in the bone marrow, a red sponge that is found in some of the bones of the body, and the body needs to consume protein, vitamins and iron to produce red blood cells and hemoglobin. Anemia is a condition in which the number of red blood cells or their hemoglobin level is lower than normal. In anemia, the destruction of red blood cells is faster than the replacement of new blood cells, so the blood will have less red blood cells, resulting in reduced oxygen transport to body tissues and a feeling of fatigue. Anemia or anemia means taken without blood, which is the quantitative or qualitative defect of hemoglobin, a molecule within the red blood cells. Because hemoglobin carries oxygen from the lungs to the body tissues, anemia can lead to hypoxia or hypoxia in tissues and organs. Because the cells of the body depend on oxygen for their survival, different degrees of anemia can create different clinical conditions. Three main causes of anemia include excessive blood loss (acute or severe bleeding and chronic but slowly but over a long period of time), excessive red blood cell destruction (hemolysis), and weakness and inability of the hematopoietic system. (hematopoiesis). Anemia is the most common blood disease. Anemia is caused by a variety of causes. There are two main ways to classify anemia: the kinetic method based on the structure and destruction of red blood cells and the morphological method that classifies anemia based on the size of the red blood cell. The morphological method uses rapid, inexpensive, and accessible laboratory tests to determine MCV. On the other hand, early focus on the kinetic method gives the treating physician the opportunity to act faster in the presence of various causes for anemia [13].

## Conclusion

The results of this study show that the use of healthy and dietary supplement increases blood factors compared to the control group. Therefore, it can be said that the use of healthy and live drug supplements can cure anemia.

Also, based on the results and findings of this study, it can be concluded that the use of the new healthy and live drug supplement has eliminated the symptoms of ovarian cyst. The concentration of this supplement can be considered as the most appropriate therapeutic dose.

**Acknowledgment.** The authors acknowledge finical support of NBS Organic Company and Tarbiat Modares University for supporting this work.